\documentstyle[12pt]{article}
\begin{document}
\vskip 2cm
\begin{center}
{\Large \bf INTERFACE STATES IN STRESSED SEMICONDUCTOR
HETEROJUNCTION WITH ANTIFERROMAGNETIC ORDERING}\\
\vskip 1cm
{{\bf Kantser V.G., Malkova N.M.,}\\
\vskip 1cm
Institute of Applied Physics, Academy of Sciences of Moldova,\\
277028 Kishinev, Moldova,\\
{\it e-mail: malkova@lises.moldova.su}}
\end{center}

\begin{center}
{\large \bf Abstract}
\end{center}

The stressed heterojunctions with antiferromagnetic ordering  in which the
constituents have oppposite band edge symmetry and their gaps have opposite
signs have been investigated. The interface states have been shown to appear
in these heterojunctions and they are spin-split. As a result if the Fermi
level gets into one of the interface bands then it leads to magnetic ordering
in the interface plane. That is  the interface magnetization effect can
be observed.

\newpage
\section{Introduction}
The majority of semiconductor structures are known to be stressed as far as
there is a lattice mismatch of their constituents. The electron energy
spectrum of the stressed semiconductor structures is determined by the strains
besides the widths of their layers and physical parameters of the constituents.
More direct strain effect is a change of the energy spectrum which is
different in each constituent and depends upon the acoustic deformation
potentials of both conduction and valence bands.  Last time this problem has
been hard investigated in different semiconductor structures \cite{smith-90}.
On the other
hand in stressed semiconductor structures the elastic strains or their
gradients due to piezoelectric or flexometric effects can lead to static
polarization fields \cite{tagan-87}. These fields are determined by the strain
values, elastic constants, piezoelectric coefficients and other material
parameters which apparently are different in each of the alternating layers.
The polarization is known to be conditioned by the shift of the cation and
anion sublattices of the binary (or multinary) semiconductors. So it is
obvious that for different crystal growth directions of the epitaxial layer
structures the polarization vector is differently directed depending on the
crystal lattice type. For example, for the structures based on IV-VI or II-VI
semiconductors the polarization is maximum for the trigonal orientation along
[111] axis and it is directed along the same axis.

In our earlier work
\cite{kant-91} we have investigated the polarization effect on the boundary
interface states of the semiconductor heterojunction, taking into account a
specific character of this polarization influence on the semiconductor energy
spectrum (looking like the spectrum of the semiconductor undergoing the
structural phase transition with appearance of the ferroelectricity ordering).
The direct
genesis of such interface states by means of the inhomogeneous polarization
field induced by the semiconductor layer strains has been studied in
heterojunctions with both the normal bands arrangment and the inverse one
\cite{kant-91}.
In the latter structures the universal electronic interface states have been
predicted to appear. These semiconductor heterojunctions have been called
the inverted band contacts (or simpler inverted contacts) \cite{volk-85},
\cite{pank-87}. They are characterized by the opposite band edge symmetry
of their constituents and so their gaps have opposite signs. As an example of
the inverted contacts the heterojunctions based on some narrow-gap
IV-VI or
II-VI semiconductors are usually considered. It is worth noting that quite
recently in the work \cite{litv-94} the magnetic-field dependences of the
Hall coefficient in PbTe/SnTe superlattices have been interpreted
assuming that in addition to the electrons in PbTe and holes in SnTe a
third kind of charge carries appears, which have been connected with
the above-mentioned interface states.

At doping with transition or
rare-earth elements the above-mentioned semiconductors turn into dilute
magnetic ones and at low temperature they might transit to the ferromagnetic
or antiferromagnetic states. Last time the quantum structures based on such
semimagnetic semiconductors have been intensively investigated
\cite{yuan-94},\cite{yuan1-94},\cite{olver-94} because of the opportunity
their practical applications. As for the interface states in these quantum
structures the antiferromagnetic ordering with the
different signs in the initial components has been shown \cite{pank1-87}
to lead to the boundary states localized near the interface plane.

Thus the aim of this work is to study the interface states in stressed
inverted contacts based on the semimagnetic narrow-gap semiconductors with
antiferromagnetic ordering. It is worth noting that this situation might be
realized by the static spin density wave. Its co-existence with the
commensurate charge density wave in the so-called systems with electron-hole
pairing results in the spin-split and under limit doping leads to the
electronic spin ordering \cite{volk-78}. Now taking into account that the
charge density wave might be induced by the structural lattices distortions
which accompany the polarization one can affirm that the situation with
antiferromagnetic ordering will be like the one in systems with the exciton
ferromagnetism but with its specific characteristics. Firstly, the fields
of the
polarization and ferromagnetic ordering are considered to be settled. Secondly,
the main attention in the evolution of the electron spectrum will be paid to
the interface states  resulting in the interface magnetic ordering.

For a concrete definition of our calculations the heterojunctions based on
the semimagnetic narrow-gap IV-VI semiconductors will be studied.

\section{Model and spectrum of the bulk semiconductors}
Both materials of the studied heterojunctions of narrow-gap IV-VI
semiconductors are known to have a direct gap at L-points of the Brillouin
zone. So that near the gap middle there are two double degenerated bands
$L^+$ and $L^-$ with opposite coordinate symmetry. Thus the simplest model
describing the spectrum of the narrow-gap IV-VI semiconductors is the
two-band one \cite{dimm-64},\cite{mitch-66}. In the paper
\cite{pank1-87} it has been shown that in
the case of the mirror symmetry bands the energy spectrum of the stressed
semiconductor heterojunction with polarization and antiferromagnetic ordering
along the trigonal [111] crystal axis picked out as z-axis might be
described by the effective Dirac Hamiltonian
\begin{equation}
{\hat H_0}=
\left( \begin{array}{cc}
\Delta & \vec{\sigma} \vec p - i(\vec{\sigma}\vec E +L)\\
\vec{\sigma} \vec p +i(\vec{\sigma}\vec E +L) & -\Delta
\end{array} \right),
\end{equation}
where $\vec{\sigma}=(\sigma_x, \sigma_y, \sigma_z)$ is the vector with the
components of the Pauli matrices, $\hat{\vec p}=-i\hbar(v_{\perp}\nabla_x,
v_{\perp}\nabla_y, v_{\parallel}\nabla_z)$, $v_{\perp, \parallel}$ being the
electron Fermi velocities, $\Delta=E_g/2$, $E_g$ is the energy gap, the vector
$\vec E$ describes the polarization. Following  the results of the work
\cite{pank1-87} we introduced the scalar $L$ to describe the antiferromagnetic
ordering with the antiferromagnetic vector along z-axis. The form of this
term can be obtained using the Heisenberg like exchange Hamiltonian in
the frames of the molecular field approximations and assuming that
antiferromagnetic order can be described as a particular spin density wave
\cite{acqua-90}.
Note that the Hamiltonian $\hat{H_0}$
looks like the one which describes the energy spectrum of the exciton
ferromagnetic within the framework of the mean field approximation
\cite{volk-78}. It is quite
in  order taking into account the above mentioned analogy between these two
tasks. In the Hamiltonian (1) the upper and lower blocks are connected with
the double degenerated states $\varphi$ and $\chi$ of the conduction and
valence bands, respectively. In this work the situation with the polarization
field $\vec E$ directed along the triagonal axis [111] is considered.

After the transformation
\begin{equation}
{\hat U} =
\left( \begin{array}{cc}
i\sigma_z & 0\\
0 & 1
\end{array} \right)
\end{equation}
the Hamiltonian (1) has the form
\begin{equation}
{\hat{\tilde H_0}}={\hat{U^{-1}}\hat H_0 \hat U}=
\left( \begin{array}{cc}
\Delta & ip_z + \hat W +E\\
-ip_z + \hat W + E & -\Delta
\end{array} \right),
\end{equation}
where the operator $\hat W =\vec{\sigma}[\hat{\vec p}\vec E] + \sigma_z L$.

First of all let us consider the energy spectrum of the homogeneous
semiconductor with polarization and antiferromagnetic ordering. After simple
calculations we obtain that the energy spectrum consists of the four
spin-split energy branches:
\begin{eqnarray}
\epsilon_{1,2}^{+} & = & \sqrt{(E+W_{\pm})^2+\Delta^2+p_z^2},\nonumber\\
\epsilon_{1,2}^{-} & = & -\sqrt{(E+W_{\pm})^2+\Delta^2+p_z^2}.
\end{eqnarray}
Here $W_{\pm}=\pm \sqrt{L^2+p_{\perp}^2}$ are the eigenvalues of the operator
$\hat{W_{\pm}}$, which correspond to the eigenvectors
\begin{equation}
{\varphi^{\pm}}= \left(
\begin{array}{c}
1\\
\frac{p_y-ip_x}{L + W_{\pm}}
\end{array} \right) \varphi_0^{\pm},
\end{equation}
where $\varphi_0^{\pm}$ is a normalized factor.
The branches $\epsilon_{1,2}^+$ and $\epsilon_{1,2}^-$ describe two spin-split
conduction and valence bands, respectively.
Taking into account the form of the wave functions for the average value of
the spin one gets
\begin{eqnarray}
\vec{S}_{1,2}^{+} & = & \frac{4|\epsilon_{1,2}^{+}|}{|E\pm\sqrt{p_{\perp}^2+
L^2}-
\epsilon_{1,2}^{+}|}\frac{1}{L \pm \sqrt{L^2+p_{\perp}^2}}(p_y,-p_x,L)
\nonumber\\
\vec{S}_{1,2}^{-} & = & \frac{4|\epsilon_{1,2}^{-}|}{|E\pm\sqrt{p_{\perp}^2+
L^2}-
\epsilon_{1,2}^{-}|}\frac{1}{L \pm \sqrt{L^2+p_{\perp}^2}}(p_y,-p_x,L).
\end{eqnarray}
So one can see that polarization and antifferromagnetic ordering split the
Kramer's spin degeneracy. Each of the branches of the conduction
$\epsilon_{1,2}^{+}$ or the valence $\epsilon_{1,2}^-$ bands are characterized
by the opposite directions of the average spin value $\vec S$. As it follows
from (6) $\vec S$ is directed along the vector
\begin{equation}
\vec I= L \vec n + [\vec n \vec p_{\perp}],
\end{equation}
where $\vec n$ is an unit vector along z-axis, $\vec p_{\perp}=(p_x,p_y,0)$.

\section{Interface states of the stressed inverted contact}
Now let us consider as inhomogeneous semiconductor structure the non-symmetry
inverted contact with the axis along z-axis when besides the coordinate
dependence of the band gap there is a coordinate dependence of the
polarization field, the parameter of the antifferomagnetic ordering being
constant in both semiconductors. The positions of  gap centres of the
constituents are different in non-symmetry inverted contact, so the
Hamiltonian must include a coordinate depending work-function $V(z)$.
To simplify analytical calculation we put that the gap-function $\Delta(z)$,
the polarization function $E(z)$ and the work-function $V(z)$ all are
determined by a single function of $z$ so that
\begin{equation}
\Delta(z)=\Delta_0 f(z), E(z)=E_0 f(z), V(z)=V_0f(z),
\end{equation}
where apparently in the inverted contact the signs of the asymptotics
$f(z\rightarrow \pm \infty)$ are opposite, $\Delta_0, E_0, V_0$ are constants.
Two different cases may be
considered: 1.~$f(+\infty)>0, f(-\infty)<0$; and 2.~$f(+\infty)<0, f(-\infty)
>0$.

So the Hamiltonian of the system is
\begin{equation}
{\hat H}= \left(
\begin{array}{cc}
\Delta-V & ip_z+ \hat W +E\\
-ip_z + \hat W +E & -\Delta + V
\end{array}
\right).
\end{equation}
By means of the unitary transformation
\begin{equation}
{\hat V}=\left(
\begin{array}{cc}
\cos{\Theta} & -\sin{\Theta}\\
\sin{\Theta} & \cos{\Theta}
\end{array}
\right),
\end{equation}
where the angle $\Theta$ satisfies the condition
\begin{equation}
\Delta_0 \cos {2\Theta}-E_0 \sin {2\Theta} + V_0=0,
\end{equation}
the Hamiltonian $\hat H$ is transformed to
\begin{displaymath}
{\hat{\tilde H}} = \hat{V^{-1}}\hat H \hat V =
\end{displaymath}
\begin{equation}
\left(
\begin{array}{cc}
-W^{\pm}\sin{2\Theta} & -\sqrt{E^2+\Delta^2-V^2}+W^{\pm}\cos{2\Theta}+ip_z\\
-\sqrt{E^2+\Delta^2-V^2}+W^{\pm}\cos{2\Theta}-ip_z & 2V+W^{\pm}\sin{2\Theta}
\end{array}
\right).
\end{equation}
{}From (12) we immediately obtain that the Schrodinger equation
\begin{equation}
(\hat{\tilde H}-\epsilon)
\left(
\begin{array}{c}
\tilde{\varphi}^{\pm}\\
\tilde{\chi}^{\pm}
\end{array}
\right)=0,
\end{equation}
where
\begin{displaymath}
\left(
\begin{array}{c}
\tilde{\varphi}^{\pm}\\
\tilde{\chi}^{\pm}
\end{array}
\right)=\hat{U}^{-1} \left(
\begin{array}{c}
\varphi^{\pm}\\
\chi^{\pm}
\end{array}
\right),
\end{displaymath}
has a solution with $\tilde{\chi}^{\pm}=0$. This is a zero-mode. It is worth
noting that the same states for diferent particular cases have been obtained
in the papers \cite{kant-91},
\cite{volk-85}, \cite{pank-87} by means of a supersymmetry quantum mechanics
and they in its term  have been called Weyl states.

In the case when $f(+\infty)>0$ and $f(-\infty)<0$ there is the following
solution of the equation (12)
\begin{equation}
\epsilon_i^{\pm}=\mp \frac{E_0V_0-\Delta_0\sqrt{E_0^2+\Delta_0^2-V_0^2}}
{\Delta_0^2+E_0^2}\sqrt{p_{\perp}^2+L^2}.
\end{equation}
The function $\tilde{\varphi}^{\pm}$
satisfies the equation
\begin{equation}
(ip_z+W^{\pm}(z))\tilde{\varphi}^{\pm}=0,
\end{equation}
where
\begin{displaymath}
W^{\pm}(z)=\sqrt{E_0^2+\Delta_0^2-V_0^2}\left( f(z)\pm \sqrt{p_{\perp}^2+L^2}
\frac{\Delta_0V_0+E_0\sqrt{E_0^2+\Delta_0^2-V_0^2}}
{(\Delta_0^2+E_0^2)\sqrt{E_0^2+\Delta_0^2-V_0^2}}\right).
\end{displaymath}
This function plays the same role as the superpotential in the supersymmetry
quantum mechanics method \cite{kant-91},\cite{pank-87}. From (15) one can
see that the states
$\epsilon_{i}^{\pm}$ are of the inteface type because the function
$\tilde{\varphi}^{\pm}$ is localized at the interface boundary.
At the given asymptotics of the $f(z)$ function
the wave functions $\tilde{\varphi}^{\pm}$ are normalized under the conditions
\begin{equation}
\sqrt{p_{\perp}^2+L^2}<\frac{(\Delta_0^2+E_0^2)\sqrt{E_0^2+\Delta_0^2-V_0^2}}
{\Delta_0V_0+E_0\sqrt{E_0^2+\Delta_0^2-V_0^2}}.
\end{equation}
So the interface states (14) are restricted both in the energy space and in
the momentum space.

In the case
of the opposite asymptotics the interface solutions are described by the same
expressions (14)-(16) by replacing $\Delta_0\rightarrow -\Delta_0$,
$p_z\rightarrow -p_z$.

\section{Conclusions}
Comparing these interface states with those of the stressed semiconductor
heterojunction without the antiferromagnetic ordering \cite{kant-91} one can
see that in this situation the spectrum of the interface states is not
linear in $p_{\perp}$. Moreover in contrast to the interface states arising
in the simple inverted contact \cite{volk-85} or in the homogeneous
semiconductor with antiferromagnetic domain wall \cite{pank-87} in the case
of the stressed inverted contact with antiferromagnetic ordering there is
a gap between  the electron-like and
the hole-like states, which is determined by the parameter of the
antiferromagnetic ordering L.

Each interface state
\begin{displaymath}
\Psi^{\pm}= \left(
\begin{array}{c}
\tilde{\varphi}^{\pm}\\
0
\end{array}
\right)
\end{displaymath}
is nondegenerated and the average spin value, for example for the first type
of the asymptotics, is
\begin{equation}
<\Psi^{\pm}|\Sigma|\Psi^{\pm}>=C exp\left( -\frac{2}{\hbar v_{\parallel}}
\int W^{\pm}(z) dz \right) \frac{2}{L \pm \sqrt{p_{\perp}^2+L^2}}
(p_y,-p_x,L),
\end{equation}
where C is a constant which is determined by a normolize condition.
After averaging over the electron momentum $p_{\perp}$ for the taken
symmetrical spectrum model one gets
\begin{equation}
<\vec{S}^{\pm}> \sim \pm (\sqrt{L^2+p_{\perp max}^2}-L)(0,0,L),
\end{equation}
where $p_{\perp max}$ is defined by the condition (16).
That is the average spins of the $\Psi^{+}$ and
$\Psi^{-}$ states are opposite directed along z-axis.

As it follows from (14) when $f(+\infty)>0, f(-\infty)<0$ under the condition
$\Delta_0^2>V_0^2$ the energy level $\epsilon_{i}^+$ is situated higher than
$\epsilon_{i}^-$ while under the condition $V_0^2>\Delta_0^2$ they switch
their positions. So the state with the average spin down is higher than the
state with the spin up. For another asymptotics: $f(+\infty)<0, f(-\infty)>0$,
the state $\epsilon_i^-$ with the spin down is higher than the state
$\epsilon_i^+$ with the spin up.

Comparing the expression (4) for the energy levels of the homogeneous
semiconductors and (14) for interface heterojunction states one gets that the
interface states are situated nearer  to the middle of the gap of the
constituents. Thus if in the studied semiconductor heterojunctions the Fermi
level, for example by means of doping, gets into one of the two-dimensional
interface bands,
then it leads to the magnetic ordering into the interface plane. The magnetic
moment as it follows from (15) is exponential attenuated moving away from
the interface plane.

Figures 1 and 2 show the rough sketch of the interface energy spectrum arising
in the stressed inverted contact with antiferromagnetic ordering for both
types of the functions $f(z)$ asymptotics. Solid lines correspond to the bulk
semiconductor bands while dashed lines to the interface states. Arrows show
the average spin direction. Note that in accordance with the taken geometry
of the studied heterojunctions the energy branches of the constituents are
the same but their spin directions are opposite in the initials
semiconductors.

The interface magnetization effect investigated in this work can be observed
also in the normal semiconductor heterojunction when the gaps of the initial
semiconductors have the same signs. But in this situation as it have been
shown \cite{kant-91} for stressed semiconductor heterostructure
the interface
states appear inside either the bulk
valence or conduction bands of the original semiconductos, and they are
restricted in the momentum space. So in this case the effect of the interface
magnetization might take place under more rigorous conditions.
\section{Acknowledgement}
We would like to thank Prof. M.Kriechbaum for useful discussions.

\newpage
\begin{center}
{\bf LIST CAPTIONS\\
for the paper Kantser V.G., Malkova N.M. "Interface states in stressed
semiconductor heterojunction with antiferromagnetic ordering"}
\end{center}
Fig. 1 Rough sketch of the interface energy spectrum in the stressed
inverted
contact with antiferromagnetic ordering  for the asymptotics $f(+\infty)>0,
f(-\infty)<0$ ($\Delta_0^2>V_0^2$).
Solid lines show the energy branches of the constituents and dashed lines
show the interface states. The arrows show the average spin direction.\\
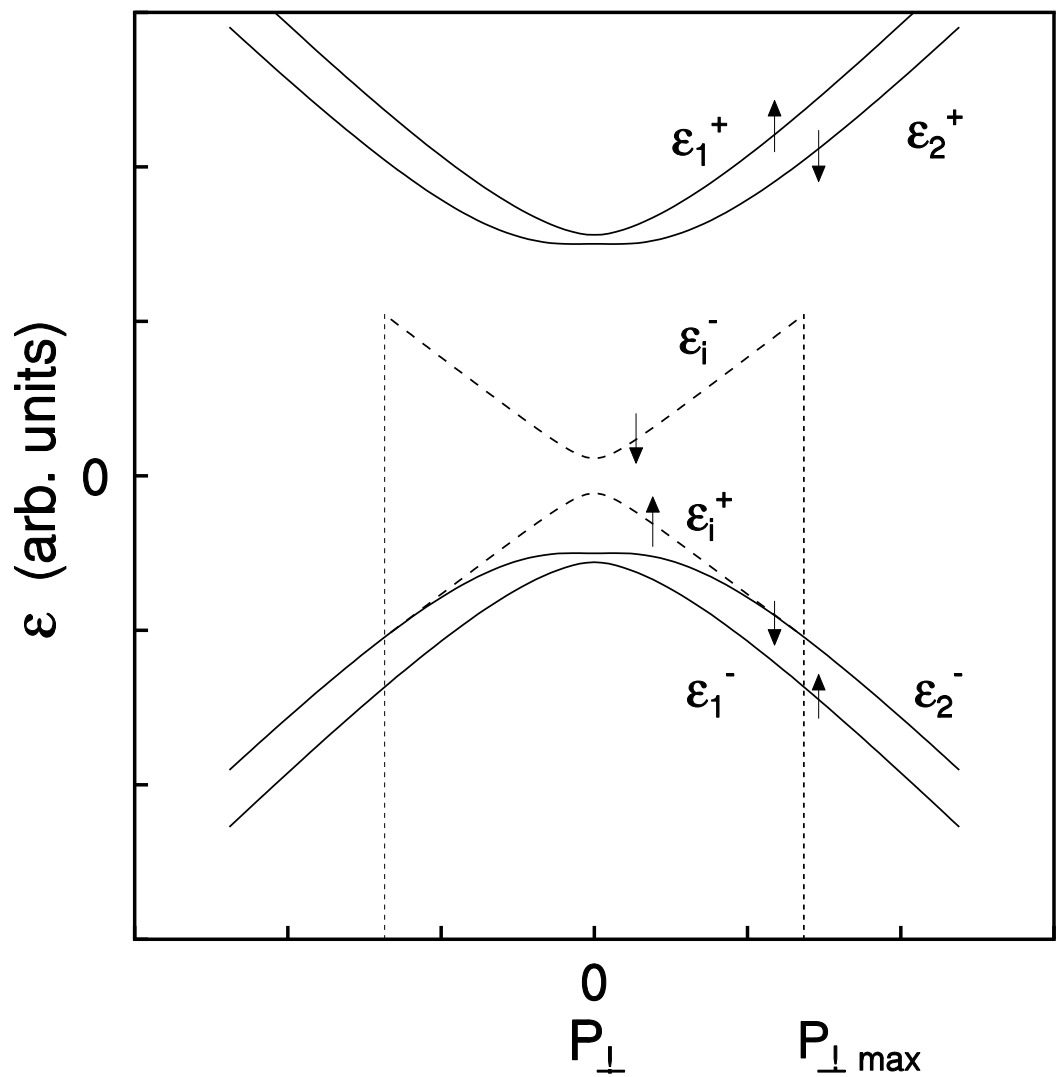
Fig. 2 The same as in the fig.1 but for the asymptotics
$f(+\infty)<0, f(-\infty)>0$.

\end{document}